\documentclass{aa}
\usepackage{graphicx}
\begin{document}
   \title{Multiwavelength studies of the Seyfert 1 galaxy NGC7469. II - X-ray and UV observations with XMM-Newton}

   \author{A. J. Blustin
          \inst{1}
          \and
          G. Branduardi-Raymont
          \inst{1}
          \and
          E. Behar
          \inst{2}
          \and
          J. S. Kaastra
	  \inst{3}
          \and
          G. A. Kriss
	  \inst{4}
          \and
          M. J. Page
          \inst{1}
          \and
	  S. M. Kahn
	  \inst{5}
          \and
	  M. Sako
	  \inst{6}
	  \and
	  K. C. Steenbrugge
	  \inst{3}
						  }

   \offprints{A. J. Blustin\\
             \email{ajb@mssl.ucl.ac.uk}}

   \institute{MSSL, University College London,
             Holmbury St. Mary, Dorking, Surrey RH5 6NT, England
             \and
             Physics Department, Technion, Haifa 32000, Israel
             \and
             SRON National Institute for Space Research, Sorbonnelaan 2, 3584 CA Utrecht, Netherlands
             \and
	     Space Telescope Science Institute, 3800 San Martin Drive, Baltimore, MD 21218, USA
             \and
             Department of Physics, Columbia University, 550 West 120th Street, New York, NY 10027, USA
             \and
	     Caltech, Pasadena, CA 91125, USA
	  }

   \date{Received 23 December 2002 / Accepted 13 February 2003}

   \abstract{We present an XMM-Newton observation of NGC 7469, including studies of the X-ray and UV variability, 0.2$-$10~keV spectral continuum, Fe K$\alpha$ emission line and the first-ever high-resolution X-ray spectrum of the soft X-ray warm absorber. We compare the properties of this X-ray warm absorber with the UV warm absorber as seen in a FUSE observation one year previously.
The 0.2$-$10~keV spectral continuum is best fitted by a power-law plus two blackbody model. An Fe K$\alpha$ emission line is visible which consists of a single narrow component and is well-modelled by a simple gaussian.
Narrow absorption and emission lines in the soft X-ray RGS spectrum demonstrate the existence of a multi-phase warm absorber with a range in log $\xi$ of $\sim$ 2 to $\sim$ $-$2 where $\xi$ is in erg cm s$^{\rm -1}$. The warm absorber is blueshifted by several hundred km s$^{\rm -1}$. The highest-ionisation phase of the absorber is the best constrained and has an overall equivalent Hydrogen column of order 10$^{\rm 20}$ cm$^{\rm -2}$; we find that its ionisation parameter is consistent with that of the warm emitter which generates the narrow emission lines. We identify this high ionisation absorber with the low-velocity phase of the UV absorber observed by FUSE.
   }

   \maketitle
%
%________________________________________________________________

\section{Introduction}

The Seyfert 1.2 galaxy NGC 7469 (z=0.0164; de Vaucouleurs et al. \cite{devaucouleurs}) was discovered by Forman et al. (\cite{forman}) to be an X-ray emitter using the UHURU satellite. Since then it has been extensively studied by numerous X-ray missions and found to have a complex and variable X-ray spectrum. A soft excess over the power-law continuum was first detected using EXOSAT by Barr (\cite{barr}) and has subsequently been studied using the Einstein Observatory (Turner et al. \cite{turner1991}), ROSAT (Turner et al. \cite{turner1993} and Brandt et al. \cite{brandt}), ASCA (Guainazzi et al. \cite{guainazzi}) and most recently BeppoSAX (De Rosa et al. \cite{derosa}). Although Barr (\cite{barr}) found the soft excess to remain constant whilst the hard band varied, the opposite situation has been observed by Guainazzi et al. (\cite{guainazzi}), who saw significant variability in the soft excess but little change in the hard power-law. Reynolds \cite{reynolds}, however, does not detect a soft excess in the ASCA data.

The most recent work on the X-ray and UV variability of NGC 7469 by Nandra et al. (\cite{nandra2000}), using a $\sim$ 30 day RXTE observation, found strong spectral variability from one day to another, and suggests that the X-ray power-law index $\Gamma$ is correlated with the UV flux, that the broadband X-ray flux is correlated with the level of the UV continuum, and that the UV variability is correlated with the extrapolation of the X-ray power-law to soft X-ray/EUV wavelengths. According to Nandra et al. (\cite{nandra2000}) this implies that the X-ray emission arises from comptonisation of UV photons. The lengths of the time lags between variations in different UV bands observed by HST and IUE (Kriss et al. \cite{kriss}) support the idea that at least part of the UV emission of the active nucleus of NGC 7469 originates from the reprocessing of high-energy radiation. Collier et al. (\cite{collier}) draw the same conclusion from time lags observed in optical data. The UV variability itself is small on short timescales: Welsh et al. (\cite{welsh}) found, using HST, that the variability over the course of an hour is at or below the Poisson noise level. Overall, they saw a 4\% change in the UV continuum (1315~${\rm \AA}$) flux during their $\sim$ 11.5 hour observation, and did not observe any time lags between different UV bands. Papadakis et al. (\cite{papadakis}), using the same RXTE dataset as Nandra et al. (\cite{nandra2000}), find a frequency-dependent time lag between hard and soft X-rays. 

Evidence for hard X-ray reflection in NGC 7469 was discovered by Piro et al. (\cite{piro}) using GINGA; a reflection hump was observed as well as an iron fluorescence line at 6.4~keV. ASCA data have indicated the presence of a narrow Fe K$\alpha$ line (Guainazzi et al. \cite{guainazzi}) or both narrow and broad components (Reynolds \cite{reynolds}), or a relativistically broadened discline (Nandra et al. \cite{nandra1997}). The existence of both broad and narrow components to the Fe K$\alpha$ emission line has been supported by BeppoSAX data (De Rosa et al. \cite{derosa}), which also contain evidence of a reflection hump and a high energy cut-off of the underlying power-law at 150~keV.

Various authors have also searched for signs of a warm absorber in the soft X-ray spectrum. Evidence for this was found in ASCA data by Reynolds (\cite{reynolds}), George et al. (\cite{george}) and Kriss et al. (\cite{kriss}), who fit \ion{O}{vii} and \ion{O}{viii} edges to the soft X-ray spectrum. More recent data from BeppoSAX (De Rosa et al. \cite{derosa}) did not require a warm absorber.

Our $\sim$ 40 ks XMM-Newton observation of NGC 7469 provides the first high-resolution soft X-ray spectrum of this source and offers the best opportunity yet to look for the spectral signatures of a warm absorber. Since the advent of high resolution astronomical X-ray spectroscopy with XMM-Newton and Chandra, it has been known that the main manifestation of warm absorption in Seyfert X-ray spectra is narrow absorption lines (Kaastra et al. \cite{kaastra2000}), as has long been observed in the UV spectra of AGN. The XMM-Newton observation of NGC 7469 was carried out one year after a UV spectrum of this object was obtained with FUSE (described in a companion paper by Kriss et al. \cite{kriss2003}), and this provides an excellent opportunity to compare the results of the analysis in the two wavebands and to see whether this can take us any further in understanding the circumnuclear environment of NGC 7469.

%__________________________________________________________________

\section{Observations and data analysis}

NGC 7469 was observed by XMM-Newton, in RGS Consortium Guaranteed Time, on 26th December 2000 for a total of about 40 ks. The observation was carried out in two halves, the first slightly longer than the second, with an approximately 7.5 ks gap between them. The instrument modes and observation durations are listed in Table~\ref{modes}.

%__________________________________________________ One column table
   \begin{table*}
    
      \caption[]{Instrument modes and exposure times.}
         \label{modes}
     $$
         \begin{array}{p{1.5in}p{2in}p{1.5in}}
            \hline
            \noalign{\smallskip}
            Instrument      &  Mode & Exposure time (seconds) \\
            \noalign{\smallskip}
            \hline
            \noalign{\smallskip}
            EPIC-MOS 1 & Large Window (medium filter) & 41388    \\
            EPIC-MOS 2 & Full Frame (medium filter) & 41388       \\
            EPIC-PN & Small Window (medium filter) & 40698  \\
            RGS 1 & Spectroscopy & 41467             \\
            RGS 2 & Spectroscopy & 41464             \\
            OM & Image (UVW2 filter) & 28500            \\
            \noalign{\smallskip}
            \hline
         \end{array}
          $$   
  \end{table*}

\begin{figure}
  \centering
   \includegraphics[width=9cm]{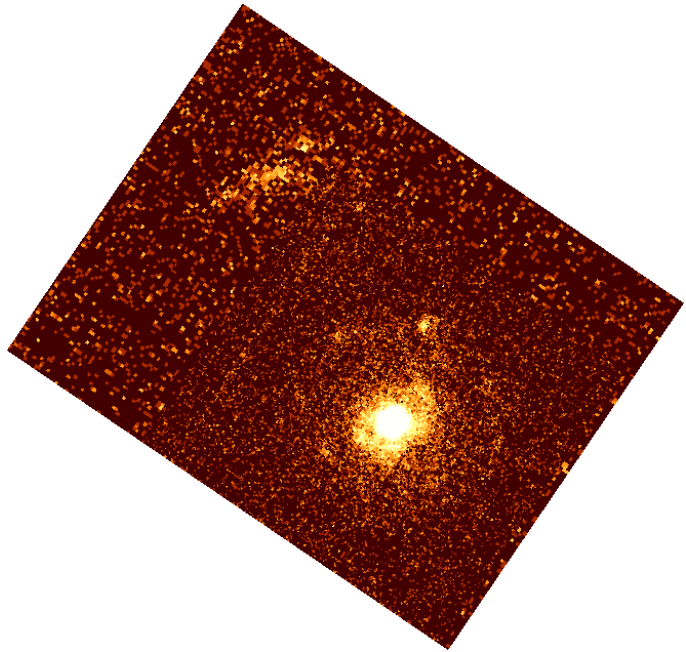}
\caption{UV image of NGC 7469 (bottom right) and IC 5283 (top left). North is up, East to the left; image size approx. 140\arcsec x 110\arcsec. The nuclear image of NGC 7469 is saturated in order to show the spiral arms. Note that NGC 7469 is within an imaging window with 0.5\arcsec \, pixels, whereas IC 5283 is in a 1\arcsec \, pixel region. }   
\label{uv_image}
\end{figure}

The OM (Mason et al. \cite{mason}) was operated in Rudi-5 imaging mode during our observation. This gave a series of 30 UVW2 filter (180$-$225~nm) images of the galaxy at 2.3\arcsec \, angular resolution, which were combined using FTOOLS to form the image in Fig.~\ref{uv_image}. This image shows UV emission from both NGC 7469 (bottom right) and IC 5283 (top left), 1.3\arcmin \, away. The UV emission of NGC 7469 is dominated by the active nucleus and central starburst (see e.g. Scoville et al. \cite{scoville}; these two components are unresolvable with the OM), although the inner spiral arms, about 30\arcsec \, across, and some much fainter outer arms, up to 60\arcsec \, in diameter, are also visible. The UV emission in IC 5283 appears more homogeneous, although three distinct sections are visible.

The long series of images also allowed us to produce a UV lightcurve for the AGN. They were processed using the \verb/omichain/ routine under SAS V5.3, and photometry of the active nucleus was performed with GAIA (the Graphical Astronomy and Image Analysis tool produced by the Central Laboratory of the Research Councils, UK) using an aperture of 10\arcsec \, diameter. Background subtraction was done using counts extracted from an annular region around the source, avoiding the spiral arms, and deadtime and coincidence loss corrections were applied to the count rates. For purposes of comparison with the UV FUSE observation, we also obtained a flux point with the same area aperture as that of FUSE (30\arcsec \, square): log $\nu$ F$_{\rm \nu}$ = $-$10.02 $\pm$ 0.04 (where $\nu$ F$_{\rm \nu}$ is in erg cm$^{\rm -2}$ s$^{\rm -1}$) at 2120 ${\rm \AA}$.

The EPIC data were processed under SAS V5.3.3. The pn and MOS spectra were extracted using the \verb/especget/ task with source regions of 45\arcsec \, radius to avoid the extraction region extending beyond the edge of the pn data window. The MOS data were found to be piled up, so only single events were used; for the pn spectra we used both single and double events. An X-ray lightcurve was obtained from the pn data (since the MOS was affected by pile-up) using \verb/xmmselect/ with a 45\arcsec \, source radius. In each case, a background spectrum (or lightcurve) was extracted using the same size region as for the source. Both X-ray and UV lightcurves were analysed using the XRONOS tools, and the EPIC spectra were analysed in XSPEC. 

The RGS (den Herder et al. \cite{denherder}) spectra were extracted with \verb/rgsproc/ under SAS V5.3.3. Background-subtraction is performed with the SAS using regions adjacent to those containing the source in the spatial and spectral domains. Since SAS V5.3.3 automatically corrects the response matrices for effective area differences between RGS1 and RGS2, it was possible to combine the spectra channel by channel. About one tenth of the RGS data was unusable due to calibration problems. The resulting combined RGS spectrum was analysed in SPEX 2.00 (Kaastra et al. \cite{kaastra2002a}).
%__________________________________________________________________

\section{Lightcurves}

\begin{figure}
   \centering
   \includegraphics[width=9cm]{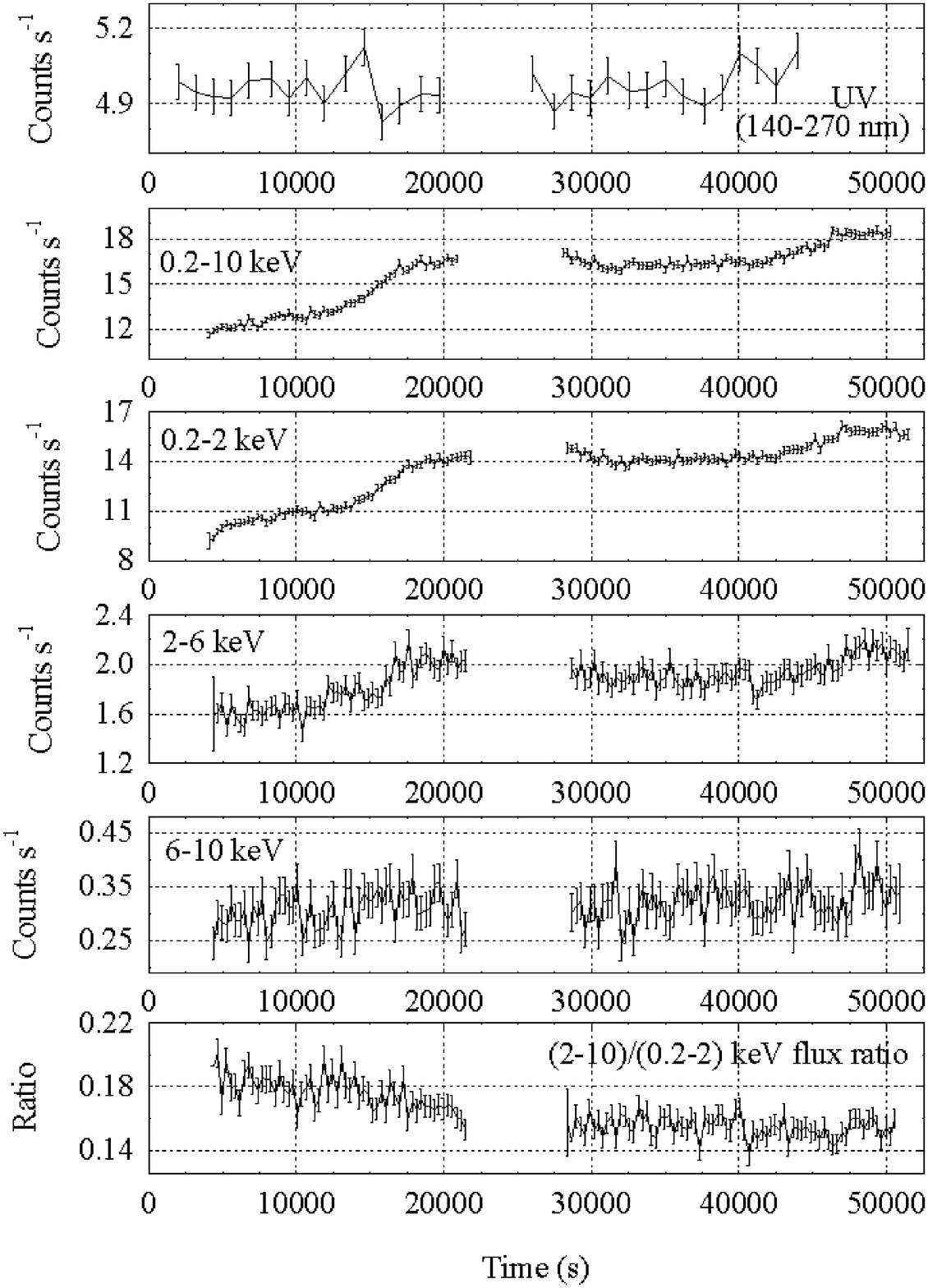}
      \caption{Multiwavelength background-subtracted lightcurves: top, UV (140$-$270~nm) lightcurve from the OM, in 1000 s time bins; middle four, pn lightcurves in the bands 0.2$-$10, 0.2$-$2, 2$-$6 and 6$-$10~keV, in 300 s time bins; bottom, the hardness ratio (2$-$10~keV / 0.2$-$2~keV flux) in 300 s time bins.
              }
         \label{ltcv}
   \end{figure}

The 0.2$-$10~keV EPIC-pn lightcurve of NGC 7469 is shown in Fig.~\ref{ltcv}; there is an overall $\sim$ 50\% increase in flux over the course of the observation in this band. The fastest significant change, at around 12000 s into the observation, is from 14 counts s$^{\rm -1}$ to 16 counts s$^{\rm -1}$ - a 14\% change - in 2 ks. Using light-crossing time arguments, this implies an emitting region of less than or equal to 4~AU.

\begin{figure}
   \centering
   \includegraphics[angle=-90,width=8cm]{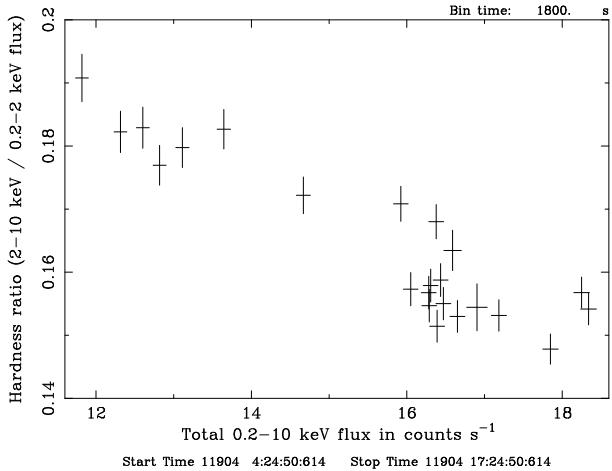}
      \caption{Hardness -- intensity plot in 30 minute time bins.
              }
         \label{colour}
   \end{figure}

Fig.~\ref{ltcv} also contains the lightcurves in the different X-ray bands (from the pn) and UV (from the OM). It indicates that the source was most variable at soft X-ray energies during our observation, with the 6$-$10~keV lightcurve showing only a gradual increase. The hardness ratio in the bottom panel of Fig.~\ref{ltcv} shows that the source gets softer as its flux increases. This is strongly supported by the hardness -- intensity plot in Fig.~\ref{colour}; the correlation coefficient for the datapoints is 0.92, with a probability of obtaining this level of correlation from a random distribution of $\sim$ 10$^{\rm -10}$. The UV data show no significant variability within the 10\% errors (consistent with the Welsh et al. (\cite{welsh}) results), whilst the X-ray bands are far more changeable; the least variable band (6$-$10~keV) increased by $\sim$ 20\% over the 45 ks observation, whereas the most variable band (0.2$-$2~keV) increased by 55\% in the first 15 ks.

It is clear from these lightcurves that there has been significant spectral variability during this observation, with the spectrum being softer in the second part of the observation than in the first. 

%-------------------------------------------------------------
%
%_____________________________________________________________

%-------------------------------------------------------------
%
%_____________________________________________________________

%__________________________________________________________________

\section{Continuum and Fe K$_\alpha$ line}

Since the MOS data were found to be heavily piled-up, the X-ray continuum fitting was done using pn data. The 0.2$-$10~keV pn spectrum for the whole observation was binned to 250 counts per bin, to provide 50 eV bins in the region of the Fe K$\alpha$ line. The best fit to this spectrum was found to be a power-law plus two blackbodies to represent the soft excess, absorbed by the ISM of our Galaxy, with the Fe K$\alpha$ emission line modelled as a single narrow gaussian (the parameters are given in Table~\ref{2bb}); however, there were still significant residuals at low energies, and above about 7.5 keV. Comptonisation (\emph{comptt} in Xspec) and reflection (\emph{pexriv} in Xspec and \emph{refl} in SPEX) models were also tried, although the $\chi$$^{\rm 2}$ was worse in both cases; our primary aim in modelling the EPIC-pn spectrum is to obtain a good estimate of the shape of the continuum underlying the RGS spectrum, so the (perhaps unphysical) power-law plus two blackbody model is used as a convenient parameterisation.

%_______________________________________________
   \begin{table*}
    
      \caption[]{Fits to the 0.2$-$10~keV X-ray continuum, the Fe K$\alpha$ emission line region and the 0.32$-$2.07~keV RGS continuum.}
         \label{2bb}
     $$
         \begin{array}{p{0.7in}p{1.2in}p{1.2in}p{1.1in}p{1.3in}p{0.85in}}
            \hline
            \noalign{\smallskip}
            Parameter & Total PN spectrum  & Hard PN spectrum & Soft PN spectrum & Combined PN, MOS1 and MOS2 spectrum & RGS spectrum \\
            \noalign{\smallskip}
            \hline
            \noalign{\smallskip}
            N$_{\rm H Gal}$$^{\mathrm{a}}$           & 4.82 & 4.82 & 4.82 & 4.82 & 4.82 \\
            $\Gamma$$^{\mathrm{b}}$                  & 1.749 $\pm$ 0.005 & 1.732 $\pm$ 0.007 & 1.755 $\pm$ 0.007 & 1.7 $\pm$ 0.1 & 1.749$^{\mathrm{q}}$ \\
            N$_{\rm P}$$^{\mathrm{c}}$               & 7.28 $\pm$ 0.03 & 6.80 $\pm$ 0.04 & 7.52 $\pm$ 0.04 & 6 $\pm$ 1 & 6.3 $\pm$ 0.5 \\
            kT$_{\rm BB1}$$^{\mathrm{d}}$            & 0.112 $\pm$ 0.001 & 0.114 $\pm$ 0.001 & 0.111 $\pm$ 0.001 & -- & 0.112$^{\mathrm{q}}$ \\
            N$_{\rm BB1}$$^{\mathrm{e}}$             & 2.66 $\pm$ 0.02 & 2.14 $\pm$ 0.03 & 3.07 $\pm$ 0.03 & -- & 1.95 $\pm$ 0.07 \\
            kT$_{\rm BB2}$$^{\mathrm{f}}$            & 0.303 $\pm$ 0.004 & 0.304 $\pm$ 0.007 & 0.303 $\pm$ 0.004 & -- & 0.303$^{\mathrm{q}}$ \\
            N$_{\rm BB2}$$^{\mathrm{g}}$             & 7.0 $\pm$ 0.1 & 5.6 $\pm$ 0.1 & 8.1 $\pm$ 0.1 & -- & 8 $\pm$ 2 \\
            E$^{\mathrm{h}}$                         & 6.40 $\pm$ 0.02 & 6.39 $\pm$ 0.02 & 6.41 $\pm$ 0.03 & 6.40 $\pm$ 0.02 & -- \\
            FWHM$^{\mathrm{i}}$                      & 130 $^{\rm +80}_{\rm -110}$ & 60 $^{\rm +100}_{\rm -60}$ & 150 $^{\rm +2500}_{\rm -150}$ & 160 $\pm$ 70 & -- \\
            V$_{\rm broad}$$^{\mathrm{j}}$           & 6000 $\pm$ 5000 & 3000 $^{\rm +5000}_{\rm -3000}$ & 7000 $^{\rm +100000}_{\rm -7000}$ & 8000 $\pm$ 3000 & -- \\
            cz$^{\mathrm{k}}$                        & -300 $\pm$ 900 & 100 $\pm$ 1000 & -1000 $\pm$ 2000 & -500 $\pm$ 800 & -- \\
            F$^{\mathrm{l}}$                         & 2.7 $\pm$ 0.5 & 3.2 $\pm$ 0.6 & 2.4 $\pm$ 0.7 & 3.0 $\pm$ 0.4 & -- \\
            $\chi$$^{\rm 2}_{\rm red}$$^{\mathrm{m}}$ & 1.80$^{\mathrm{n}}$ (399) & 1.12$^{\mathrm{n}}$ (607) & 1.28$^{\mathrm{n}}$ (692) & 1.25$^{\mathrm{o}}$ (61) & 1.17$^{\mathrm{p}}$ (994) \\
            \noalign{\smallskip}
            \hline
         \end{array}
          $$   
\begin{list}{}{}
\item[$^{\mathrm{a}}$] Galactic absorbing column in 10$^{\rm 20}$ cm$^{\rm -2}$, fixed (Elvis et al. \cite{elvis})
\item[$^{\mathrm{b}}$] photon index
\item[$^{\mathrm{c}}$] power-law normalisation in 10$^{\rm -3}$ photons keV$^{\rm -1}$ cm$^{\rm -2}$ s$^{\rm -1}$ at 1 keV
\item[$^{\mathrm{d}}$] temperature of blackbody 1 in keV
\item[$^{\mathrm{e}}$] normalisation of blackbody 1; 10$^{\rm -4}$ L$_{39}$/ D$_{10}^{2}$, where L$_{39}$ is the source luminosity in units of 10$^{39}$ ergs s$^{\rm -1}$ and D$_{\rm 10}$ is the distance to the source in units of 10 kpc
\item[$^{\mathrm{f}}$] temperature of blackbody 2 in keV
\item[$^{\mathrm{g}}$] normalisation of blackbody 2; 10$^{\rm -5}$ L$_{39}$/ D$_{10}^{2}$, where L$_{39}$ is the source luminosity in units of 10$^{39}$ ergs s$^{\rm -1}$ and D$_{\rm 10}$ is the distance to the source in units of 10 kpc
\item[$^{\mathrm{h}}$] measured rest frame energy of the Fe K${\alpha}$ line in keV
\item[$^{\mathrm{i}}$] FWHM of Fe K${\alpha}$ line in eV
\item[$^{\mathrm{j}}$] velocity broadening (FWHM) of Fe K${\alpha}$ line in km s$^{\rm -1}$
\item[$^{\mathrm{k}}$] redshifted velocity of Fe K${\alpha}$ line in km s$^{\rm -1}$ (negative values indicate blueshifts)
\item[$^{\mathrm{l}}$] flux of Fe K$\alpha$ line in 10$^{\rm -5}$ photons cm$^{\rm -2}$ s$^{\rm -1}$
\item[$^{\mathrm{m}}$] reduced $\chi$$^{\rm 2}$ of the total fit; the degrees of freedom are given in brackets
\item[$^{\mathrm{n}}$] in the range 0.2$-$10~keV
\item[$^{\mathrm{o}}$] in the range 5$-$8~keV
\item[$^{\mathrm{p}}$] in the range 0.32$-$2.07~keV
\item[$^{\mathrm{q}}$] fixed
\end{list}
  \end{table*}

\begin{figure*}
   \centering
   \includegraphics[width=18cm]{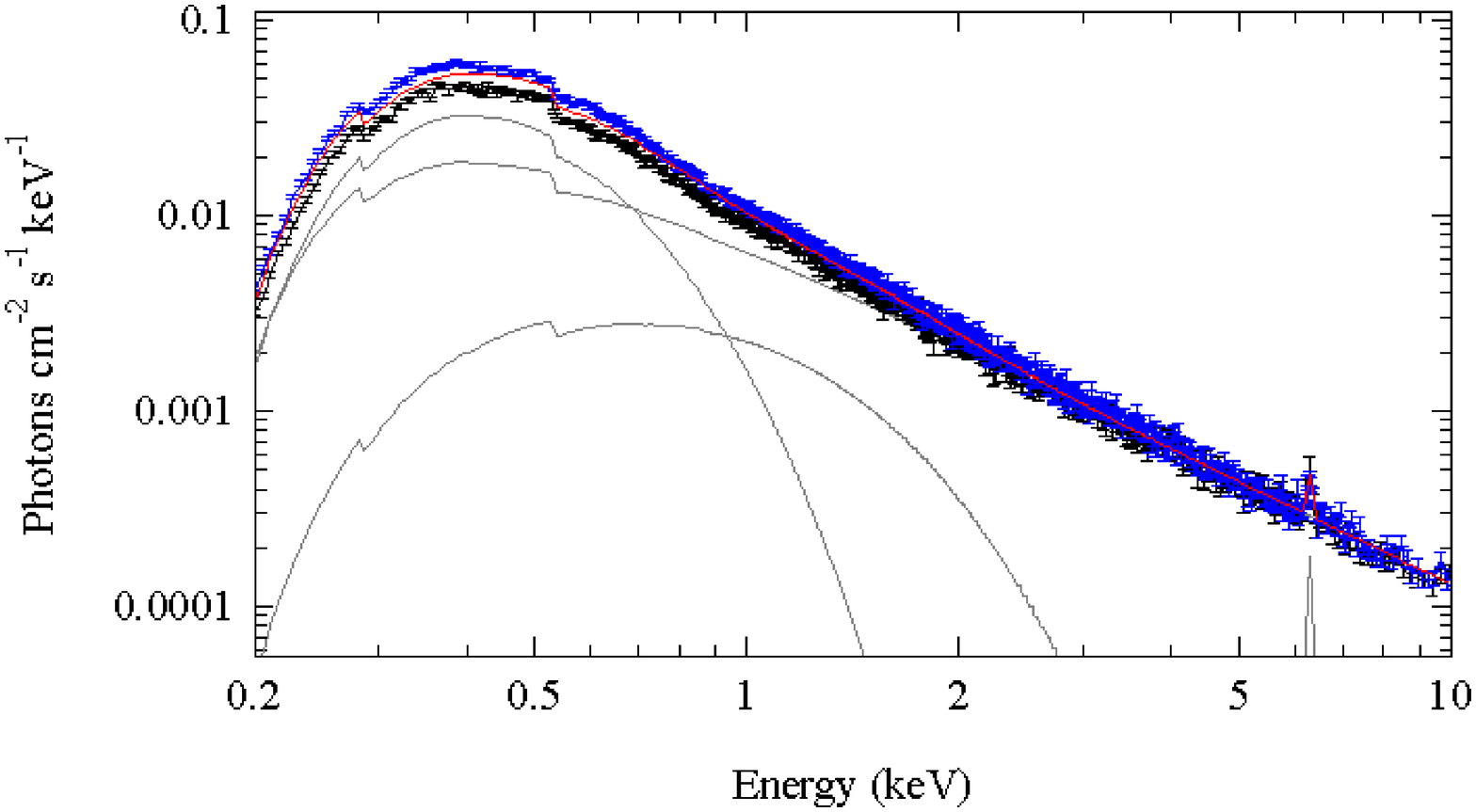}
      \caption{The soft (blue) and hard (black) PN spectra plotted with the best fit model for the total pn spectrum (red), with the two blackbodies, power-law and Fe K$\alpha$ line model components (all in grey).
              }
         \label{hard_soft}
   \end{figure*}

\begin{figure}
   \centering
   \includegraphics[width=8cm]{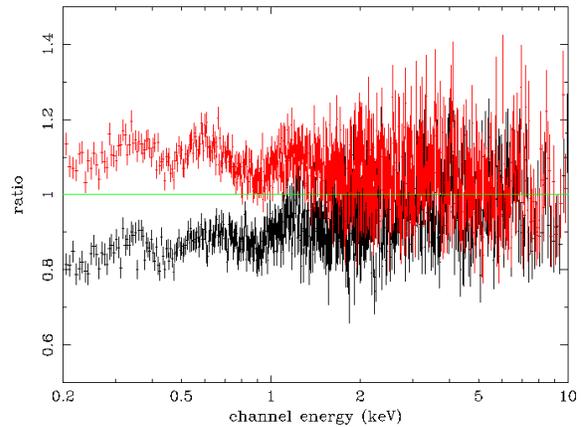}
      \caption{The ratio of the soft pn spectrum (red) and the hard pn spectrum (black) to the best fit model for the total pn spectrum. 
              }
         \label{ratio}
   \end{figure}

Because the spectrum becomes softer over the course of the observation (Fig.~\ref{ltcv}), we have also fitted continua separately to the pn spectra for each of the two parts of the observation (binned by 125 counts per bin, again to provide approximately 50 eV bins in the region of the Fe K$\alpha$ line). These two spectra are plotted alongside the model fit to the overall pn spectrum in Fig.~\ref{hard_soft}, and Fig.~\ref{ratio} shows their ratio to this model. The parameters of power-law plus two blackbody and gaussian fits (again with absorption due to our Galaxy) to the hard and soft spectra are given in Table~\ref{2bb}. Despite the increase in flux in the soft band, the shape of the soft excess does not vary significantly.

%-------------------------------------------------------------
%
%_____________________________________________________________

%-------------------------------------------------------------
%
%_____________________________________________________________

\begin{figure}
   \centering
   \includegraphics[angle=-90,width=8cm]{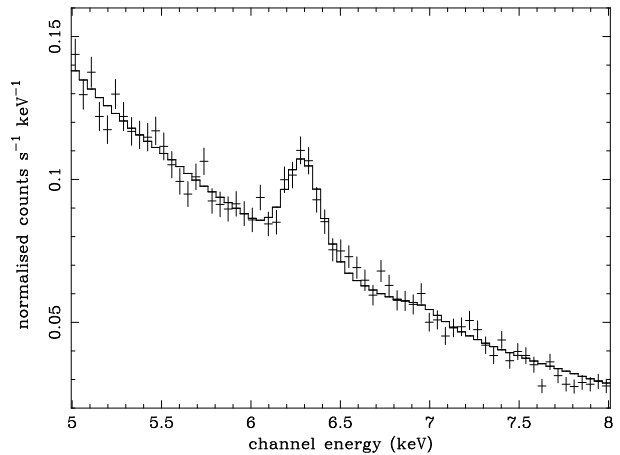}
      \caption{The (observer's frame) Fe K$\alpha$ emission line (combined MOS1, MOS2 and pn data) with the model including both Fe K$\alpha$ and Fe K$\beta$. 
              }
         \label{fe_plot}
   \end{figure}

To gain better signal-to-noise in the Fe K$\alpha$ line, we combined the total pn spectrum with the MOS1 and MOS2 spectra and binned it so as to provide 50 eV bins in the region of interest. Due to the heavy pile-up in the MOS data we could not fit a reliable continuum model to the whole combined spectrum. Instead, we fitted a power-law to the 5$-$8~keV range and used this as the underlying continuum when modelling the Fe K$\alpha$ line; the parameters are given in Table~\ref{2bb}. The line itself was best fitted by a single marginally resolved gaussian, whose parameters are listed in Table~\ref{2bb}. When a second gaussian was added in for Fe K$\beta$, with the same $\sigma$ as for K$\alpha$ and at the expected 14\% of the intensity, the reduced $\chi$$^{\rm 2}$ worsened slightly to 1.28 for 58 degrees of freedom. The resulting fit is shown in Fig.~\ref{fe_plot}. 

%-------------------------------------------------------------
%
%_____________________________________________________________

%__________________________________________________________________

\section{RGS spectrum}

\subsection{Warm absorber modelling}

We analysed our RGS data using SPEX 2.00, which contains two photoionised absorber models: \emph{slab} and \emph{xabs}. \emph{Slab} applies line and continuum absorption by individual ions to the spectral continuum, allowing the user to specify the blueshift of the medium and its turbulent velocity. \emph{Xabs} is a grid model of Xstar runs applying absorption by a column of photoionised gas at a range of different ionisation parameters, column densities, and elemental abundances. We define the ionisation parameter $\xi$ here as

   \begin{equation}
      \xi = \frac{L}{n r^{\mathrm{2}}} \,,
   \end{equation}
 
where $L$ is the source luminosity (in erg s$^{\rm -1}$), $n$ the gas density (in cm$^{\rm -3}$) and $r$ the source distance in cm, so $\xi$ has the units erg cm s$^{\rm -1}$ (Tarter et al. \cite{tarter}) which are used throughout this paper.

 \emph{Slab} and \emph{xabs} incorporate new calculations, using HULLAC (Bar-Shalom et al. \cite{bar}), for absorption due to states of M-shell iron and lower ionisation states of oxygen (\ion{O}{ii}$-$\ion{O}{vi}). \emph{Slab} is the best tool for detailed measurements of individual ion columns in a spectrum, whereas \emph{xabs} enables the physical parameters and composition of a warm absorber to be investigated directly. A more detailed description of these models is given in Kaastra et al. (\cite{kaastra2002a}).

\subsection{Continuum and absorption lines}

\begin{figure}
   \centering
   \includegraphics[width=9cm]{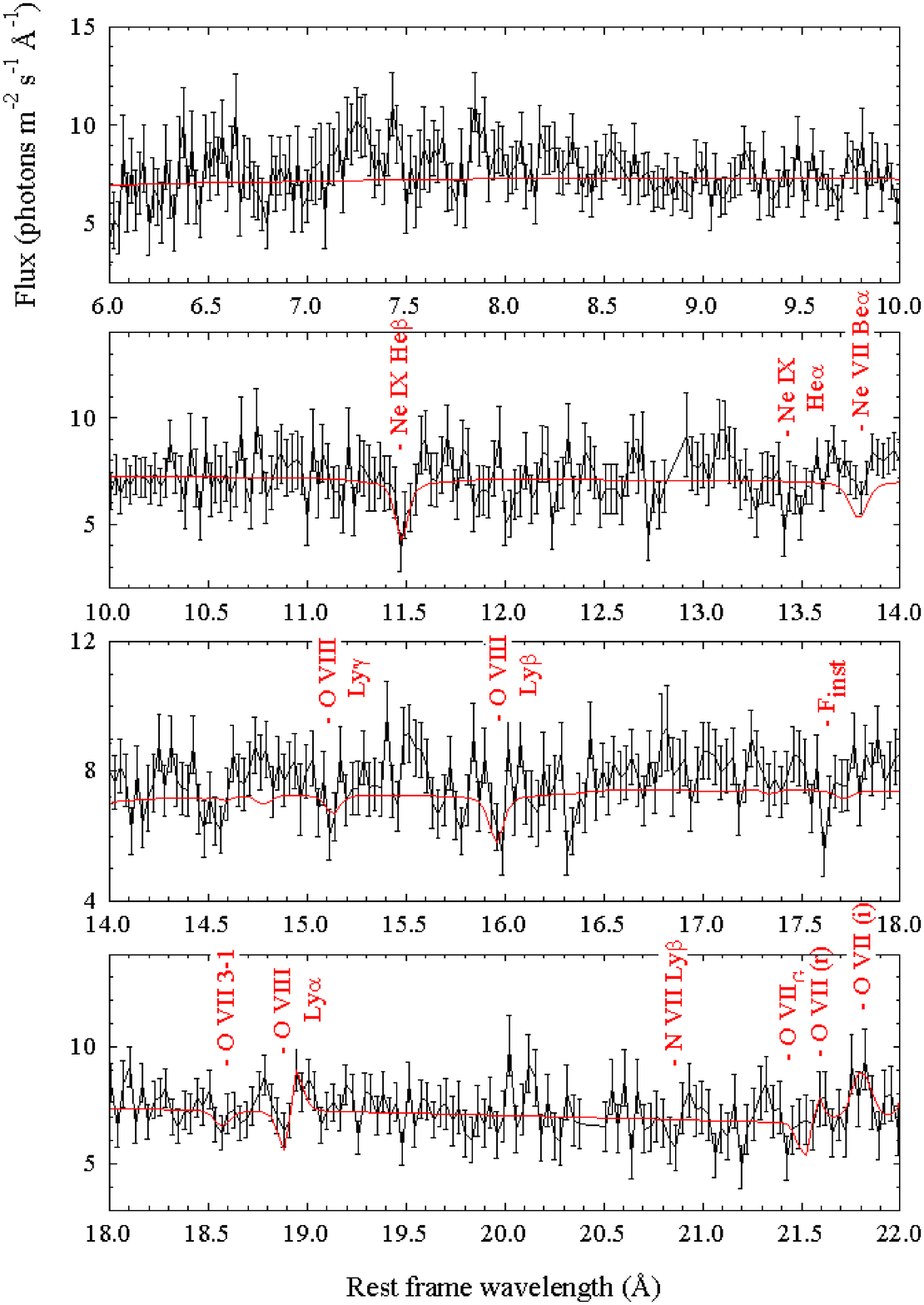}
      \caption{Combined RGS1 and RGS2 rest frame spectrum of NGC 7469 (black), binned into groups of three channels, with power-law plus two blackbody continuum model superimposed (red) including all lines listed in Table~\ref{abs_lines}, plotted over the range 6$-$22~$\rm \AA$ (0.326$-$0.564~keV). The positions of other possible spectral features are labelled. The G subscript refers to features originating in our galaxy, and F$_{\rm inst}$ is an instrumental Fluorine feature.
              }
         \label{spec_1}
   \end{figure}

\begin{figure}
   \centering
   \includegraphics[width=9cm]{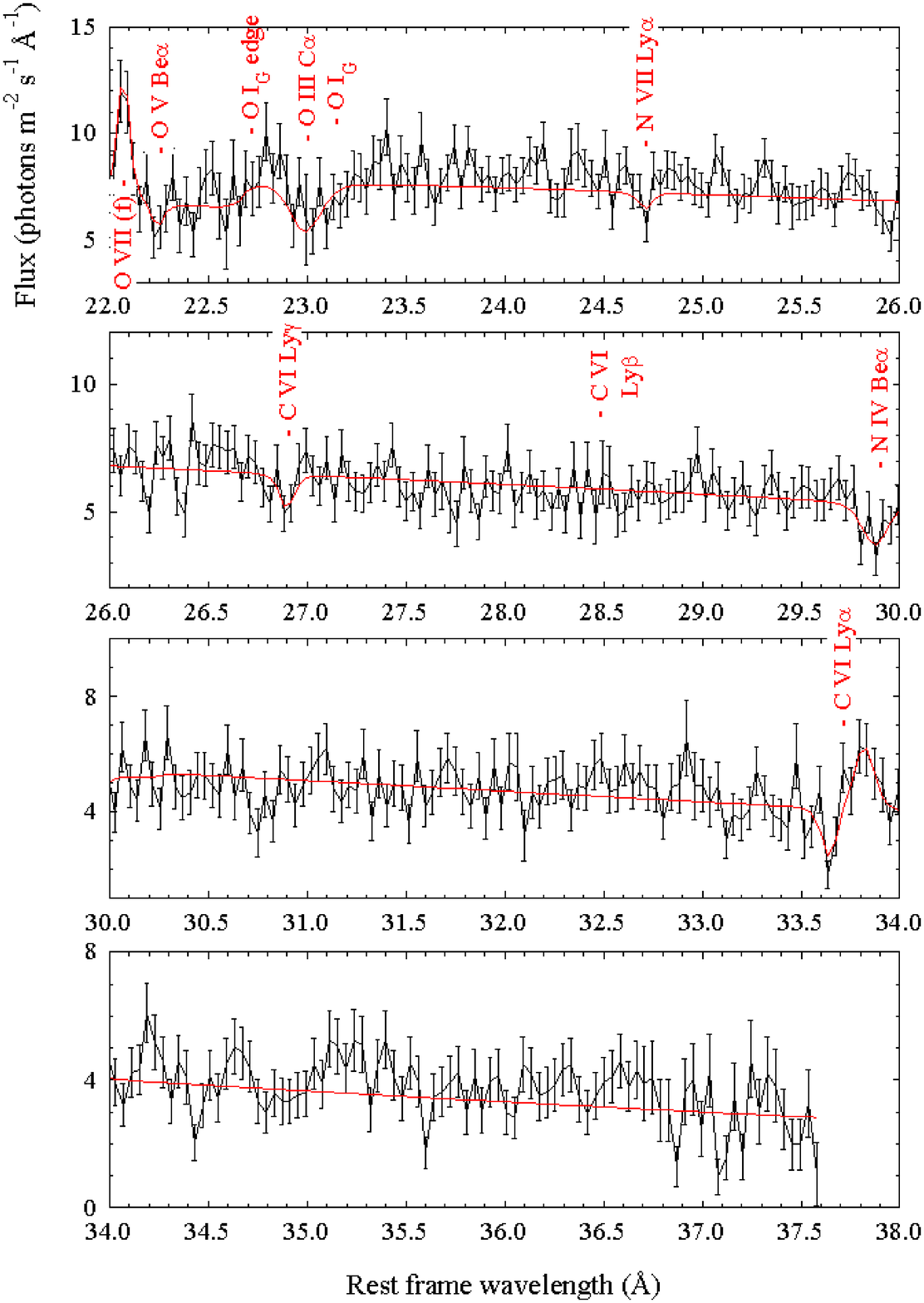}
      \caption{As for Fig.~\ref{spec_1}, plotted in the range 22$-$38~$\rm \AA$ (0.564$-$2.066~keV).
                   }
         \label{spec_2}
   \end{figure}

The RGS data were binned into groups of three channels for the analysis, slightly below the instrument resolution. The spectral continuum was represented by a Galactic absorbed power-law plus two blackbody model, using the power-law slope and blackbody temperatures derived from the pn but allowing the normalisations of the three components to vary. Problems with differences in relative normalisations between the RGS and pn were thus avoided. There is a known 0.1 difference in power-law slope between the RGS and pn; we do not take this into account in our fitting as the blackbody soft excess is by far the most important component over most of the RGS band, although it will probably have a small effect on the normalisations that we fit. We also do not take account of the warm absorber when fitting this continuum, and we discuss the accuracy of this assumption at the end of Sect. 5.3. The parameters of this continuum are given in Table~\ref{2bb}, and the RGS spectrum with the continuum overlaid is shown in Fig.~\ref{spec_1} and Fig.~\ref{spec_2}.

%-------------------------------------------------------------
%
%_____________________________________________________________
%-------------------------------------------------------------
%
%_____________________________________________________________

\begin{figure}
   \centering
   \includegraphics[width=9cm]{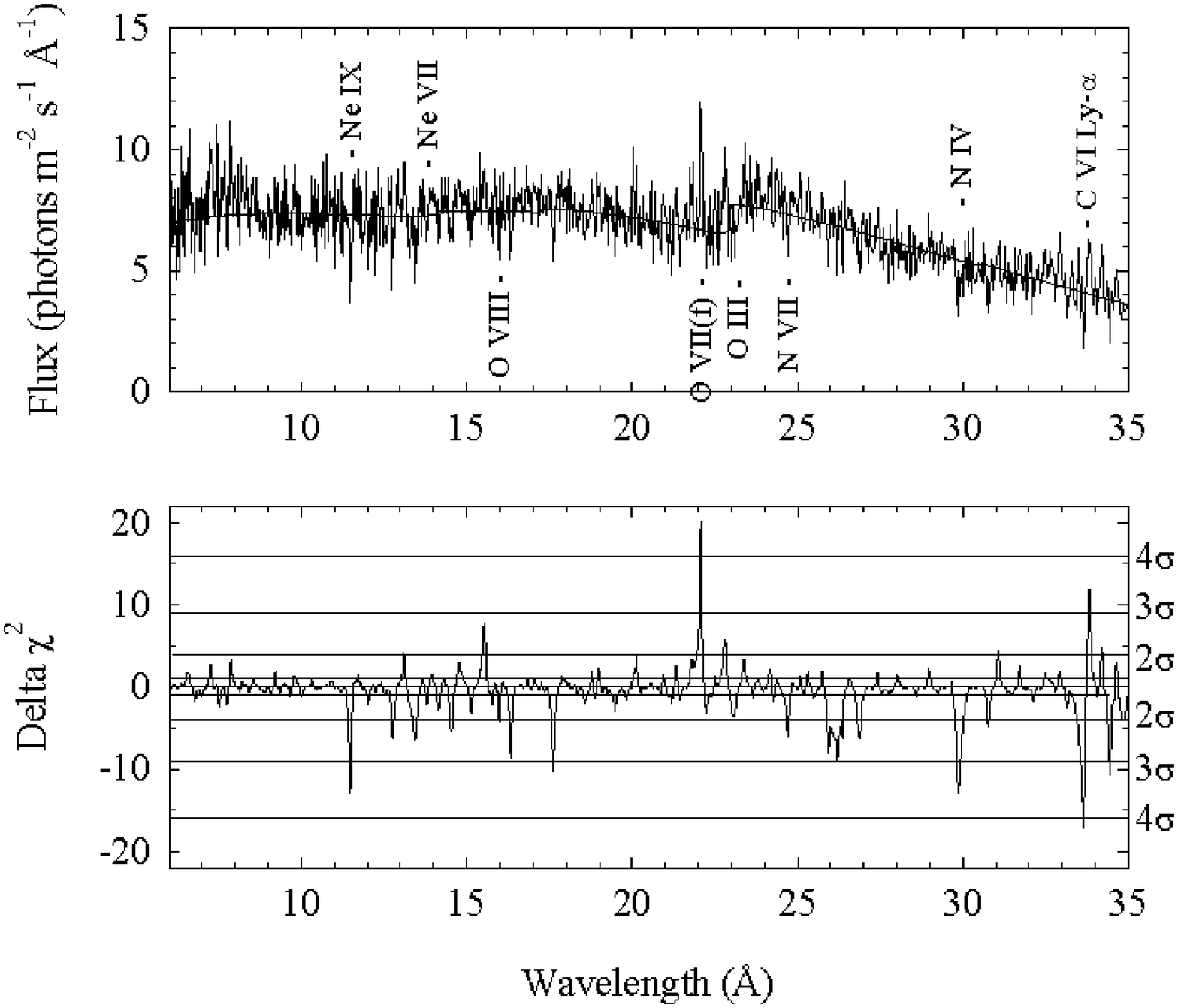}
      \caption{Top: combined RGS1 and RGS2 rest frame spectrum of NGC 7469 (6$-$38~$\rm \AA$; 0.326$-$2.066~keV) with power-law plus two blackbody continuum superimposed, and the positions of important features marked. Bottom: plot of the statistical significance of the narrow absorption and emission lines.
              }
         \label{stat_sig}
   \end{figure}

A number of narrow absorption and emission lines, unresolved at the resolution of the RGS, are visible in the RGS spectrum of NGC 7469; no intrinsic absorption edges are seen. The positions of the lines are marked in Fig.~\ref{spec_1} and Fig.~\ref{spec_2}. The statistical significance of these features was assessed by fitting a gaussian to them against a locally fitted continuum; the results of this are plotted in Fig.~\ref{stat_sig} (bottom) below a plot of the complete RGS spectrum with its power-law and two blackbody continuum model (top). Table~\ref{abs_lines} gives the identification, rest wavelength, observed wavelength, equivalent width, blueshifted velocity, flux and velocity width (where appropriate) and log $\xi$ of maximum abundance of the measured absorption and emission lines. Also given, for comparison, are the rest wavelength, blueshifted velocity, flux, velocity width and log $\xi$ of maximum abundance of the broad and narrow \ion{O}{vi} observed by FUSE (Kriss et al. \cite{kriss2003}).

%-------------------------------------------------------------
%
%_____________________________________________________________

%_____________________________________________________________
   \begin{table*}
    
      \caption[]{The rest wavelengths ($\lambda$$_{\rm rest}$), observed wavelengths ($\lambda$$_{\rm obs}$), equivalent widths (EW), blueshifts (v$_{\rm blue}$), fluxes, velocity widths (v$_{\rm broad}$, FWHM) and range of log $\xi$ (log $\xi$$_{\rm max}$) where the ion is most abundant of absorption (top section) and emission lines (bottom section) in the RGS spectrum of NGC 7469; rms errors are quoted throughout, except for log $\xi$$_{\rm max}$ where the error bars give the range where the ion has a higher relative abundance than adjacent ionisation states. Blueshifts are not given for lines whose equivalent width is consistent with zero. Also listed are rest wavelengths, blueshifts, fluxes, velocity widths and log $\xi$$_{\rm max}$ for the broad and narrow \ion{O}{vi} emission lines observed by FUSE (Kriss et al. \cite{kriss2003}).}
         \label{abs_lines}
 $$
\begin{array}{p{0.9in}p{0.6in}p{0.9in}p{0.7in}p{0.9in}p{0.7in}p{0.7in}p{0.8in}}
 \hline
 \noalign{\smallskip}
Transition & $\lambda$$_{\rm rest}$$^{\mathrm{a}}$ & $\lambda$$_{\rm obs}$$^{\mathrm{b}}$ & EW$^{\mathrm{c}}$ & v$_{\rm blue}$$^{\mathrm{d}}$ & Flux$^{\mathrm{e}}$ & v$_{\rm broad}$$^{\mathrm{f}}$ & log $\xi$$_{\rm max}$$^{\mathrm{g}}$ \\
 \noalign{\smallskip}
 \hline
 \noalign{\smallskip}
\ion{O}{viii} Ly$\alpha$ & 18.969 & 18.89 $\pm$ 0.05 & 10 $\pm$ 10    & &               & & 1.8 $\pm$ 0.3 \\
\ion{O}{viii} Ly$\beta$  & 16.006 & 15.98 $\pm$ 0.02 & 20 $\pm$ 10    & $-$600 $\pm$ 500  &               & & 1.8 $\pm$ 0.3 \\
\ion{O}{viii} Ly$\gamma$ & 15.176 & 15.12 $\pm$ 0.05 & 10 $\pm$ 10    & &               & & 1.8 $\pm$ 0.3 \\
\ion{O}{vii} 3$-$1 (r)   & 18.629 & 18.57 $\pm$ 0.04 & 10 $\pm$ 10    & &               & & 1.1 $^{\rm +0.4}_{\rm -0.5}$ \\
\ion{O}{v} Be$\alpha$    & 22.334 & 22.25 $\pm$ 0.04 & 20 $\pm$ 20    & &               & & 0$^{\rm +0.4}_{\rm -0.3}$ \\
\ion{O}{iii} C$\alpha$   & 23.058 & 22.99 $\pm$ 0.04 & 70 $\pm$ 20    & $-$800 $\pm$ 500  &               & & $-$1.8 $\pm$ 0.6 \\
\ion{N}{vii} Ly$\alpha$  & 24.781 & 24.72 $\pm$ 0.04 & 10 $\pm$ 10    & &               & & 1.5 $\pm$ 0.3 \\
\ion{N}{iv} Be$\alpha$   & 29.896 & 29.87 $\pm$ 0.03 & 60 $\pm$ 20    & $-$300 $\pm$ 400  &               & & $-$0.6 $\pm$ 0.5 \\
\ion{C}{vi} Ly$\alpha$   & 33.736 & 33.65 $\pm$ 0.03 & 50 $\pm$ 20    & $-$800 $\pm$ 200  &               & & 1.1 $^{\rm +0.4}_{\rm -0.3}$ \\
\ion{C}{vi} Ly$\gamma$   & 26.990 & 26.90 $\pm$ 0.03 & 20 $\pm$ 10    & $-$1000 $\pm$ 300 &               & & 1.1 $^{\rm +0.4}_{\rm -0.3}$ \\
\ion{Ne}{ix} He$\beta$   & 11.547 & 11.48 $\pm$ 0.02 & 50 $\pm$ 20    & $-$1600 $\pm$ 500 &               & & 1.7 $^{\rm +0.3}_{\rm -0.4}$ \\
\ion{Ne}{vii} Be$\alpha$ & 13.814$^{\mathrm{h}}$ & 13.79 $\pm$ 0.02   & 40 $\pm$ 10 & $-$500 $\pm$ 500 &  & & 0.5 $\pm$ 0.3 \\
 \noalign{\smallskip}
 \hline
 \noalign{\smallskip}
\ion{O}{viii} Ly$\alpha$ & 18.969 & 18.98 $\pm$ 0.05 & $-$20 $\pm$ 10 & 200 $\pm$ 700     & 1.4 $\pm$ 1.0 & 0$^{\rm +3000}_{\rm -0}$    & 1.8 $\pm$ 0.3\\
\ion{O}{vii} (f)         & 22.101 & 22.07 $\pm$ 0.02 & $-$80 $\pm$ 20 & $-$400 $\pm$ 200  & 6.0 $\pm$ 1.7 & 0$^{\rm +700}_{\rm -0}$     & 1.1 $^{\rm +0.4}_{\rm -0.5}$ \\
\ion{O}{vii} (i)         & 21.807 & 21.80 $\pm$ 0.04 & $-$40 $\pm$ 20 & $-$100 $\pm$ 500  & 2.7 $\pm$ 1.6 & 700$^{\rm +2000}_{\rm -700}$ & 1.1 $^{\rm +0.4}_{\rm -0.5}$ \\
\ion{O}{vi}$^{\mathrm{i, j}}$ & 1031.93 & & & 323 $\pm$ 37    & 4457 $\pm$ 223 & 4901 $\pm$ 146 & 0.5 $^{\rm +0.1}_{\rm -0.2}$ \\
\ion{O}{vi}$^{\mathrm{i, j}}$ & 1037.62 & & & 323 $\pm$ 37    & 2241 $\pm$ 115 & 4901 $\pm$ 146 & 0.5 $^{\rm +0.1}_{\rm -0.2}$ \\
\ion{O}{vi}$^{\mathrm{i, k}}$ & 1031.93 & & & $-$163 $\pm$ 55 & 1221 $\pm$ 151 & 1061 $\pm$ 82 & 0.5 $^{\rm +0.1}_{\rm -0.2}$ \\
\ion{O}{vi}$^{\mathrm{i, k}}$ & 1037.62 & & & $-$163 $\pm$ 55 & 616 $\pm$ 78   & 1061 $\pm$ 82 & 0.5 $^{\rm +0.1}_{\rm -0.2}$ \\
\ion{C}{vi} Ly$\alpha$  & 33.736 & 33.82 $\pm$ 0.03 & $-$90 $\pm$ 30 &  700 $\pm$ 200    & 3.6 $\pm$ 1.3 & 700$^{\rm +800}_{\rm -700}$ & 1.1 $^{\rm +0.4}_{\rm -0.3}$ \\
 \noalign{\smallskip}
 \hline
 \end{array}
          $$   
\begin{list}{}{}
\item[$^{\mathrm{a}}$] rest wavelength in ${\rm \AA}$
\item[$^{\mathrm{b}}$] observed wavelength in ${\rm \AA}$
\item[$^{\mathrm{c}}$] equivalent width in m${\rm \AA}$
\item[$^{\mathrm{d}}$] blueshifted velocity in km s$^{\rm -1}$
\item[$^{\mathrm{e}}$] observed flux of line in 10$^{\rm -5}$ photons cm$^{\rm -2}$ s$^{\rm -1}$
\item[$^{\mathrm{f}}$] velocity broadening (FWHM) in km s$^{\rm -1}$
\item[$^{\mathrm{g}}$] log $\xi$ where the ion is at its maximum relative abundance; $\xi$ has the units erg cm s$^{\rm -1}$, and the error bars give the range where the ion has a higher relative abundance than adjacent ionisation states
\item[$^{\mathrm{h}}$] from Behar \& Netzer (2002)
\item[$^{\mathrm{i}}$] measured in the UV by FUSE (Kriss et al. \cite{kriss2003})
\item[$^{\mathrm{j}}$] from the Broad Line Region
\item[$^{\mathrm{k}}$] from the Narrow Line Region
\end{list}
  \end{table*}
%_____________________________________________________________

As can be seen in Fig.~\ref{spec_1}, the $\alpha$ transitions of \ion{O}{viii}, \ion{O}{vii} and \ion{Ne}{ix} are not as deep as the corresponding $\beta$ transitions. This appears to be because the deepest resonance absorption lines can be accompanied by re-emission, as shown by the P-Cygni profile of \ion{C}{vi} Ly$\alpha$.

\begin{figure}
   \centering
   \includegraphics[width=9cm]{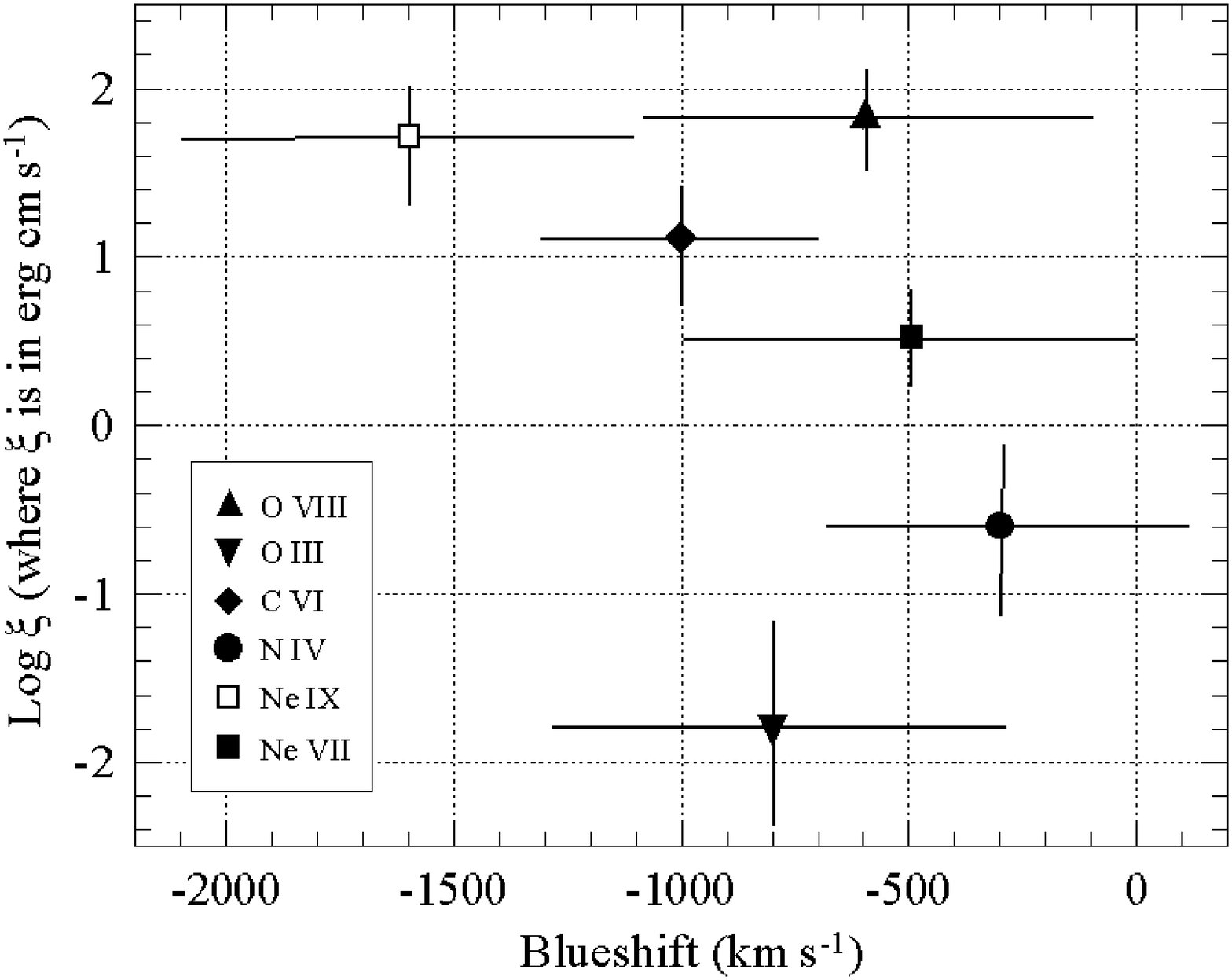}
      \caption{Blueshift plotted against log $\xi$ of maximum abundance for ions observed in the RGS spectrum of NGC 7469.
              }
         \label{xil_vblu}
   \end{figure}

The blueshifts of these absorption lines (except \ion{O}{viii} Ly$\alpha$ and \ion{C}{vi} Ly$\alpha$, for which the apparent blueshifts are affected by re-emission, and those lines with equivalent widths consistent with zero) are plotted against log $\xi$ of maximum abundance for their ionisation state in Fig.~\ref{xil_vblu}, which shows that the ions in the absorber originate from a very wide range of ionisation levels. The warm absorber is blueshifted at every level of ionisation; the weighted average of the blueshift of all the absorption lines is $-$800 $\pm$ 100 km s$^{\rm -1}$. Unfortunately, there are not enough measurable ions to be certain as to whether there is any significant variation in blueshift with ionisation level. As a very crude measure, the weighted average of blueshifts from the highest ionisation phase (which appears to cluster around log $\xi$ = 1.5) is $-$900 $\pm$ 100 km s$^{\rm -1}$, whilst the weighted average of blueshifts from the lower ionisation gas (log $\xi$ from 0.5 downwards) is $-$600 $\pm$ 200 km s$^{\rm -1}$.

%-------------------------------------------------------------
%
%___________________________________________

If we assume that all of the observed \ion{O}{viii} and \ion{O}{vii} come from the high-ionisation phase, then we can use the ratio of the absorbed fluxes of \ion{O}{viii} Ly$\beta$ and the \ion{O}{vii} 3$-$1 resonance line to estimate its ionisation parameter (\ion{O}{viii} Ly$\alpha$ and \ion{O}{vii} 2$-$1 (r) are partially filled in by re-emission and so cannot reliably be used for this purpose). Log $\xi$ for the high ionisation phase can be estimated in this way to be 1.6 $^{\rm +0.7}_{\rm -0.4}$. 

There is no unambiguous evidence for iron absorption or emission in the RGS spectrum of NGC 7469; this is in strong contrast to the warm absorbers of other Seyferts observed so far by XMM-Newton and Chandra (cf NGC 3783 (Kaspi et al. \cite{kaspi2000}), NGC 5548 (Kaastra et al. \cite{kaastra2002b})). The apparent lack of iron absorption is probably due to the low overall column of the absorber. If log $\xi$ of the high ionisation phase is 1.6, then the \emph{xabs} model in SPEX 2.00 can be used to fit an overall equivalent hydrogen column, assuming a turbulent velocity of 100 km s$^{\rm -1}$, of (1.5 $\pm$ 0.9) x 10$^{\rm 20}$ cm$^{\rm -2}$. At this column, iron absorption would not be significant at the signal-to-noise of our spectrum. Although the velocity width of the UV absorption was found to be rather lower than what we assume for the X-ray absorber, at 25 km s$^{\rm -1}$, it actually makes no difference to the column obtained given the resolution and signal-to-noise of the RGS spectrum.

\subsection{Emission lines}

The most important emission lines in our spectrum are those of \ion{C}{vi} Ly$\alpha$ (33.736 ${\rm \AA}$ in the rest frame; rest frame wavelengths used throughout), \ion{O}{vii} forbidden (22.101 ${\rm \AA}$) and intercombination (21.807 ${\rm \AA}$) lines and \ion{O}{viii} Ly$\alpha$ (18.969 ${\rm \AA}$). Table~\ref{abs_lines} lists the parameters of these lines. They were fitted without taking into account any neighbouring absorption, and thus the blueshifts of the lines may be underpredicted. Although none of the lines were significantly resolved by the RGS, it is possible that their true widths could also be concealed by the absorption.

The third member of the \ion{O}{vii} emission triplet, the resonance line, is not unambiguously present. However, we observe \ion{O}{vii} 3$-$1 resonance absorption, which means that there must also be absorption from \ion{O}{vii} 2$-$1 (r). Because no \ion{O}{vii} 2$-$1 (r) absorption is actually observed in the expected position, it must be being filled in by \ion{O}{vii} 2$-$1 (r) emission - the \ion{O}{vii} (r) emission and absorption, for the 2$-$1 transition, are cancelling each other out. A similar situation exists with \ion{O}{viii}, where \ion{O}{viii} Ly$\alpha$ is far less deep than the corresponding Ly$\beta$ or Ly$\gamma$ absorption lines, and is being partially filled in by the adjacent \ion{O}{viii} Ly$\alpha$ emission line.

We can estimate the depth and blueshift of \ion{O}{vii} 2$-$1 (r) - and \ion{O}{viii} Ly$\alpha$ - absorption lines from fitting the depth and blueshift of lines higher up in the series (applying absorption by \ion{O}{vii} and \ion{O}{viii} ions using the \emph{slab} model in SPEX). If we also assume that the blueshifts and velocity widths of \ion{O}{vii} (r) and \ion{O}{viii} Ly$\alpha$ emission lines are the same as that of the \ion{O}{vii} forbidden emission line, which is the most accurately measured, then we can estimate their fluxes. 

The ratio of these fluxes (after Porquet \& Dubau \cite{porquet}),

   \begin{equation}
      X_{\rm ion} = \frac{u}{w} \,,
   \end{equation}

where u and w are the fluxes of the \ion{O}{viii} Ly$\alpha$ and \ion{O}{vii} (r) emission lines respectively, is indicative of the ionisation parameter of the emitter. Once we have estimated the ionisation parameter of the emitter we can compare it with that of the absorber.

Using \emph{slab} in SPEX 2.00, we fitted absorbing columns of \ion{O}{vii} and \ion{O}{viii} (again assuming a velocity width of 100 km s$^{\rm -1}$) to the \ion{O}{vii} 3$-$1 (r), \ion{O}{viii} Ly$\beta$ and \ion{O}{viii} Ly$\gamma$ absorption lines, obtaining log N$_{\rm \ion{O}{vii}}$ = 16.5 $^{\rm +0.4}_{\rm -1.9}$ and log N$_{\rm \ion{O}{viii}}$ = 17.3 $\pm$ 0.3 (where N is in cm$^{\rm -2}$), at an average blueshift of $-$900 $\pm$ 400 km s$^{\rm -1}$. The ratio of these columns implies a log $\xi$ of 2.1 $^{\rm +1.6}_{\rm -0.5}$, consistent with the value obtained in Section 5.2 using the straightforward ratio of the fluxes absorbed by \ion{O}{vii} 3$-$1 (r) and \ion{O}{viii} Ly$\beta$. 

We applied \ion{O}{vii} and \ion{O}{viii} absorption with these fitted columns and blueshift to the spectrum using the \emph{slab} model in SPEX. Assuming that the blueshift of the emitter is $-$400 km s$^{\rm -1}$, and that the emission lines have zero intrinsic width (taking the parameters of \ion{O}{vii} (f) as a guide), we modelled the \ion{O}{vii} (r) and \ion{O}{viii} Ly$\alpha$ emission lines with the SPEX gaussian model, and fitted their fluxes. The results of this depend upon whether we are seeing the emission lines through the warm absorber itself: if the warm absorption affects only the continuum and not the emission lines, the intrinsic flux of the emission lines can be a lot lower.

In the case where the warm absorption is only affecting the continuum, and not the emission lines, the ratio of \ion{O}{vii} (r) and \ion{O}{viii} Ly$\alpha$ emission line fluxes implies a log $\xi$ of 2.2$^{\rm +0.3}_{\rm -0.2}$. If, on the other hand, the warm absorption is applied to both emission lines and continuum, a log $\xi$ of 2.7 $\pm$ 0.2 is obtained. Both of these values are consistent with log $\xi$ of the high-ionisation phase of the absorber. The fitted fluxes of these lines, their ratios and the derived ionisation parameters are summarised in Table~\ref{em_xi}.

%_____________________________________________________________

  \begin{table}
    
      \caption[]{Estimates of the ionisation parameter of the warm emitter using the ratio of the fluxes of \ion{O}{viii} Ly$\alpha$ and \ion{O}{vii} He$\alpha$ emission lines. The fluxes were fitted assuming an \ion{O}{viii} absorbing column of 10$^{\rm 17.3}$ cm$^{\rm -2}$ and an \ion{O}{vii} column of 10$^{\rm 16.5}$ cm$^{\rm -2}$, both blueshifted by $-$900 km s$^{\rm -1}$. The \ion{O}{viii} Ly$\alpha$ and \ion{O}{vii} He$\alpha$ emission lines were assumed to have a blueshift of $-$400 km s$^{\rm -1}$ and zero intrinsic width, for fitting purposes.}
         \label{em_xi}
     $$
         \begin{array}{p{0.6in}p{0.7in}p{0.43in}}
            \hline
            \noalign{\smallskip}
             Model$^{\mathrm{a}}$ & A & B \\
             F$_{\rm \ion{O}{vii}}^{\mathrm{b}}$ & 4 $\pm$ 2 & 3 $\pm$ 1 \\
             F$_{\rm \ion{O}{viii}}^{\mathrm{b}}$ & 32 $\pm$ 6 & 5 $\pm$ 1 \\
             X$_{\rm ion}^{\mathrm{c}}$ & 7$^{\rm +8}_{\rm -3}$ & 2$^{\rm +2}_{\rm -1}$ \\
             log $\xi$$^{\mathrm{d}}$ & 2.7 $\pm$ 0.2 & 2.2$^{\rm +0.3}_{\rm -0.2}$ \\
            \noalign{\smallskip}
            \hline
         \end{array}
     $$   
\begin{list}{}{}
\item[$^{\mathrm{a}}$] A: warm absorption applied to both continuum and emission lines, B: warm absorption only applied to continuum
\item[$^{\mathrm{b}}$] model line flux in 10$^{\rm -5}$ photons cm$^{\rm -2}$ s$^{\rm -1}$, including the effects of Galactic (but not intrinsic) absorption
\item[$^{\mathrm{c}}$] ratio of fluxes of \ion{O}{viii} Ly$\alpha$ and \ion{O}{vii} (r) 
\item[$^{\mathrm{d}}$] implied log $\xi$ in erg cm s$^{\rm -1}$
\end{list}
  \end{table}

%------------------------------------------------

\begin{figure}
   \centering
   \includegraphics[width=9cm]{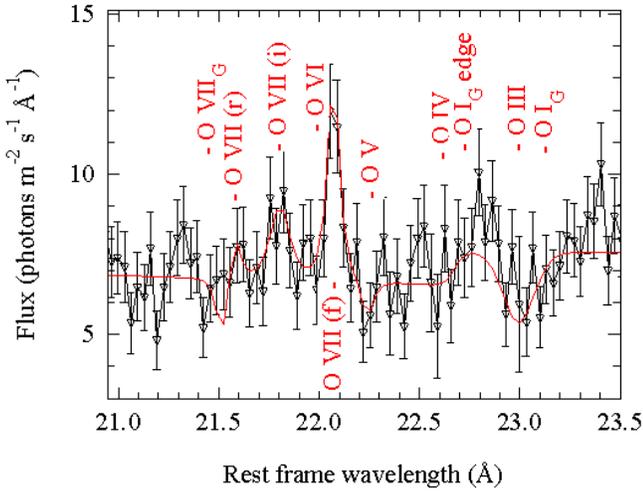}
      \caption{The \ion{O}{vii} emission line triplet region with model overposed (continuum plus fitted columns of \ion{O}{vii} and \ion{O}{viii}; the other absorption and emission lines are included with the parameters given in Table~\ref{abs_lines}). Positions of other transitions not significantly detected are also shown.
              }
         \label{o_region}
   \end{figure}

The signal-to-noise in the \ion{O}{vii} emission triplet region (Fig.~\ref{o_region}) is not very good; we can, however, use the ratio of the fluxes of the forbidden and intercombination lines to estimate an upper limit to the density of the emitter, after Porquet \& Dubau (\cite{porquet}). This ratio,

   \begin{equation}
      R = \frac{f}{i} \,,
   \end{equation}

(where f and i are the fluxes of the \ion{O}{vii} (f) and \ion{O}{vii} (i) emission lines respectively) has a value of $\sim$ 2, giving an upper limit for the density of $\sim$ 10$^{\rm 10}$ cm$^{\rm -2}$. In general, great care needs to be taken in using the Porquet \& Dubau (\cite{porquet}) diagnostic ratios for photoionised plasmas when investigating emission triplets in AGN, as they are sensitive to photoexcitation, which has been found to occur in the warm emitter of NGC 1068 (Kinkhabwala et al. \cite{kinkhabwala}).

%-------------------------------------------------------------
%
%_____________________________________________________________

What effect, then, do the warm absorber and emitter have on the fitting of the overall continuum spectrum with the EPIC-pn? The small number of absorption lines detected and the lack of absorption edges would imply that, when seen at the spectral resolution of the pn, the warm absorber would not have much of an effect on the fitting of the broad-band continuum spectrum. To test this, we used \emph{slab} to apply absorption from \ion{O}{viii}, \ion{O}{vii}, \ion{O}{v}, \ion{O}{iii}, \ion{N}{vii}, \ion{Ne}{ix} and \ion{C}{vi} to the pn spectrum (the log ion columns were estimated as 10$^{\rm 17.3}$, 10$^{\rm 16.5}$, 10$^{\rm 16.2}$, 10$^{\rm 16.6}$, 10$^{\rm 16.2}$, 10$^{\rm 17.5}$ and 10$^{\rm 17.0}$ cm$^{\rm -2}$ respectively), taking into account the differences in normalisation between the RGS and pn, and the absorption from ions not yet included in \emph{slab} (\ion{N}{iv} and \ion{Ne}{vii}) was modelled with gaussian absorption lines with the parameters given in Table~\ref{abs_lines}. The emission lines were also included with the parameters expected from warm absorber coverage of the continuum only (in the case of \ion{O}{vii} (r) and \ion{O}{viii} Ly$\alpha$) or the measured parameters from Table~\ref{abs_lines} (for \ion{O}{vii} (f), \ion{O}{vii} (i) and \ion{C}{vi} Ly$\alpha$).

We found that the inclusion of warm absorption required an increase of up to $\sim$ 10\% in the overall normalisation of the soft excess in the pn spectrum, and an increase of $\sim$ 2\% in the power-law continuum. This is obviously totally dependent on the model adopted for the warm absorption. A higher signal-to-noise spectrum would allow us to model the warm absorber more accurately and to better gauge its effect on the shape of the overall 0.2$-$10~keV continuum.

\section{Discussion and conclusions}

\subsection{Lightcurves}

The X-ray spectrum of NGC 7469 has a variable soft excess. There is no evidence in our relatively short dataset for a connection between X-ray and UV variability. Whilst there is at least 20\% variability in all of the X-ray bands, there is less than 10\% variability in the UV.

\subsection{Continuum and iron K$\alpha$ line}

We find a Galactic absorbed power-law plus two blackbody model to be a fairly good representation of the pn spectrum, although there are significant residuals at low energies which are probably instrumental in origin. The soft excess represented by the blackbodies is perhaps unlikely - due to its rapid variability - to actually be due to blackbody emission. It is more probable that the soft excess originates from comptonisation, in different parts of a highly dynamic corona at a range of temperatures, of UV flux from an accretion disc below. Our comptonisation model fits (fitting a single comptonised model to the entire soft excess; the number of variables becomes very large when more than one such model is used) are too simplistic to model the spectrum accurately. The shape of the soft excess is likely to be determined by the different populations of energetic electrons in various parts of the corona, as well as the wavelength and spatial distribution of the UV flux originating in the accretion disc. We use the blackbody models here as a convenient parameterisation to produce a correctly shaped continuum to underly the RGS spectrum.

In our highest signal-to-noise spectrum (combined pn and MOS data) the Fe K$\alpha$ emission line appears to be marginally resolved and to have a FWHM consistent with that of the \ion{O}{vi} and \ion{S}{iv} UV emission lines seen in the FUSE spectrum of NGC 7469 (Kriss et al. \cite{kriss2003}); these lines originate in the Broad Line Region. The intrinsic redshifts of these lines are consistent with that measured for Fe K$\alpha$. At any rate, the fact that Fe K$\alpha$ originates in fluorescence from cool gas might imply that any broadening would be of dynamical rather than thermal origin. The Fe K$\alpha$ emission line fitted to only the pn data (avoiding possible broadening from combining data from different instruments), though, is not broadened, and it is quite possible that the apparent width of Fe K$\alpha$ seen in the combined spectrum could be due to the presence of a weak \ion{Fe}{xxv} line. The higher spectral resolution, in this range, of the recently completed Chandra HETGS observation of NGC 7469 should constrain any broadening of this line much better.

\subsection{Warm absorber/emitter}

NGC 7469 has an outflowing warm absorber with multiple ionisation phases, as demonstrated by the presence of several narrow absorption lines from O, N, C, and Ne in the RGS spectrum. There is no significant edge absorption. Most of the ions we observe come from the highest ionisation phase, which has a log $\xi$ of 1.6$^{\rm +0.7}_{\rm -0.4}$, and an average blueshift of $-$1000 $\pm$ 500 km s$^{\rm -1}$ and an overall column of (1.5 $\pm$ 0.9) x 10$^{\rm 20}$ cm$^{\rm -2}$. This low overall column, about two orders of magnitude lower than that in the high-ionisation phase of the very deep warm absorber in NGC 3783 (Blustin et al. \cite{blustin2002}), is probably the reason why there is no unambiguous evidence of iron absorption in the spectrum; any evidence of iron could also be concealed by the low signal-to-noise in the dataset.   

We identify four narrow emission lines; the forbidden and intercombination lines from the \ion{O}{vii} triplet, \ion{O}{viii} Ly$\alpha$ and C VI Ly$\alpha$. The latter two have P-Cygni profiles and are partially absorbed by the corresponding resonance absorption lines, so their true fluxes and blueshifts, as well as any intrinsic broadening, will be obscured by the neighbouring absorption. Of the other two lines, \ion{O}{vii} (f) has the best constrained parameters and is found to have a blueshift of $-$400 $\pm$ 200 km s$^{\rm -1}$ and to be unresolved at the resolution of the RGS. Our main result from the emission lines is that they come from a gas with an ionisation parameter consistent with that of the high-ionisation phase of the warm absorber.

\subsection{Connection between the UV and X-ray absorbers} 

A UV spectrum of NGC 7469 was obtained using FUSE a year before our XMM-Newton observation, and is published in a companion paper (Kriss et al. \cite{kriss2003}). It was found that the UV absorber contains two different velocity components.

The blueshift of UV component \#1 ($-$569 km s$^{\rm -1}$) makes it the best match for the high-ionisation phase of our X-ray warm absorber (which has a weighted average blueshift of $-$900 $\pm$ 100 km s$^{\rm -1}$), taking into account the systematic wavelength uncertainty of about 100 km s$^{\rm -1}$. No transitions of \ion{O}{vi} are visible in the RGS spectrum - its column must be somewhere lower than that of \ion{O}{vii} at $\pm$ 10$^{\rm 16.5}$ cm$^{\rm -2}$. In fact, the \ion{O}{vi} columns of $\sim$ 8 x 10$^{\rm 14}$ cm$^{\rm -2}$, observed by FUSE, are far too low to be observable in our RGS spectrum. However, it is shown by Kriss et al. (\cite{kriss2003}) that the columns of \ion{O}{viii} and \ion{O}{vii} observed in the RGS spectrum, and the columns of \ion{O}{vi} and \ion{H}{i} observed in the FUSE spectrum, are all consistent with a photoionised warm absorber model with U = 6.0 and an equivalent Hydrogen column of 3.5 x 10$^{\rm 20}$ cm$^{\rm -2}$. U is defined here as the ratio of ionising photons in the Lyman continuum to the electron density; for the best-fit FUSE continuum model, $\xi$ = 24.05 x U and so U = 6.0 corresponds to log $\xi$ of 2.2. By way of comparison, the ratio of the \ion{O}{viii} and \ion{O}{vii} absorbing columns fitted to the RGS spectrum corresponds to log $\xi$ = 2.1$^{\rm +1.6}_{\rm -0.5}$, with an overall Hydrogen column of (4 $\pm$ 2) x 10$^{\rm 20}$ cm$^{\rm -2}$.

UV component \#2 has a modelled overall column of 3.8 x 10$^{\rm 18}$ cm$^{\rm -2}$ and U = 0.2 (log $\xi$ = 0.68). The systemic velocity of this phase is $-$1898 km s$^{\rm -1}$; we do not observe any ions at this blueshift and ionisation level in the RGS spectrum (see Fig.~\ref{xil_vblu}), which is unsurprising at this low overall column.

Our conclusions are, then, that the high-ionisation X-ray warm absorber is probably consistent with component \#1 of the UV warm absorber, and that there is no detected X-ray counterpart to UV component \#2 - and no detected UV counterpart to the lower-ionisation X-ray absorption. Following the conclusions of Kriss et al. (\cite{kriss2003}) on the spatial locations of the two UV absorbers, this would place the high-ionisation X-ray warm absorber (which we equate with UV \#1) inside the BLR as part of, perhaps, an accretion disc wind. The wide range of ionisation parameters observed in the X-ray absorber is also consistent with a thermally driven outflow at a larger radius, as described by Krolik \& Kriss (\cite{krolik}).

The overall picture of the circumnuclear environment is as follows: closest to the central engine is highly-ionised X-ray and UV absorbing gas, outflowing at about 600 km s$^{\rm -1}$, and outside this is the Broad Line Region, producing the broad emission lines seen in the UV and optical spectra. The soft X-ray emission lines could also originate from here. Further out still, there is UV absorbing gas, with a much lower overall column in our line of sight than the first absorber, that is flowing towards us at about 1900 km s$^{\rm -1}$. 

Taking the Wandel et al. (\cite{wandel}) reverberation measurement of the distance of the BLR from the central engine, $\sim$ 10$^{\rm 16}$ cm, and assuming log $\xi$ of 2 for the innermost absorber and a 1$-$1000 Ryd AGN luminosity of $\sim$ 10$^{\rm 44}$ erg s$^{\rm -1}$ (derived from the RGS continuum model), we can estimate a \emph{lower} limit of 10$^{\rm 10}$ cm$^{\rm -3}$ for the density of this absorber. The ratio of the fluxes of the \ion{O}{vii} (f) and \ion{O}{vii} (i) gives a strict \emph{upper} limit to the density of whichever phase produces the soft X-ray emission lines of $\sim$ 10$^{\rm 10}$ cm$^{\rm -3}$. We have shown that the ionisation state of the high-ionisation (innermost) warm absorber is consistent with that of the emitter; if we were to infer that they do originate in the same body of gas, then we could place the warm absorber/emitter coincident with the BLR. Using the overall equivalent Hydrogen column of the absorber, 4 x 10$^{\rm 20}$ cm$^{\rm -2}$, and the estimated density of 10$^{\rm 10}$ cm$^{\rm -3}$, we obtain an approximate depth for the absorber of 400 000 km. The small size of this, in comparison with the distance from the central engine, would imply a sheet-like or low filling factor structure.

Are the properties of the X-ray emission lines consistent with those of the UV lines from the BLR? Unfortunately, the intrinsic widths and blueshifts measured for the X-ray emission lines are affected by adjacent absorption, which makes the comparison difficult. The width and blueshift of \ion{O}{vii} (f), which is the strongest line and is least affected by absorption, are most similar to those of the narrow \ion{O}{vi} UV lines (Table~\ref{abs_lines}). If X-ray emission lines are formed at the distance of the Narrow Line Region, then either they are not spatially consistent with the high-$\xi$ warm absorber, or the high-$\xi$ absorber is actually located this far out and the covering factor arguments placing it within the BLR were not correct.

\begin{figure}
   \centering
   \includegraphics[width=8cm]{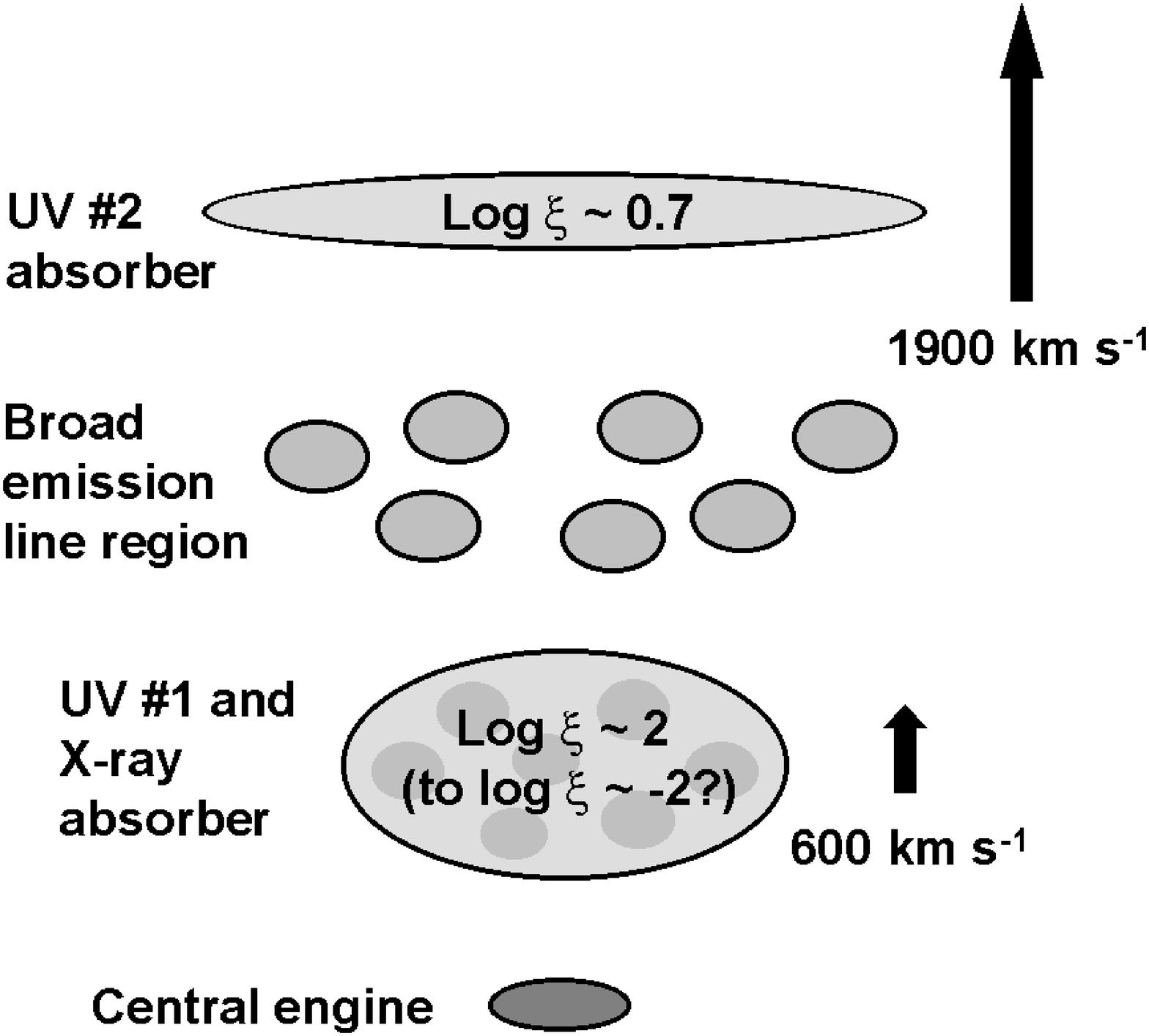}
      \caption{The overall picture of the circumnuclear region of NGC 7469 as inferred from X-ray and UV absorption and emission measurements.
              }
         \label{model}
   \end{figure}

We know much less about the low-ionisation phases of the X-ray warm absorber, as we detect far fewer ions from this regime. It is, however, clear that log $\xi$ could range down as far as -2. If we assumed that the low-ionisation X-ray absorber was spatially coincident with the high-ionisation phase, then for a range in log $\xi$ of -2 to 2 there would be a four magnitude range in density. Assuming the high-$\xi$ absorber to be at the distance of the BLR, the density would range from that of the high-ionisation line-emitting plasma at 10$^{\rm 10}$ cm$^{\rm -3}$ down to a much denser low-ionisation absorber at densities up to perhaps 10$^{\rm 14}$ cm$^{\rm -3}$. If, on the other hand, the low $\xi$ gas was assumed to have the same density as the high-ionisation absorber, it could be located up to a hundred times further out from the nucleus. Since the UV absorber may show evidence of changing outflow velocity with distance from the nucleus, and the blueshifts of the high and low $\xi$ X-ray emitter seem to be comparable, there may be cause to prefer the interpretation involving a clumpy X-ray absorbing gas with a wide density range. Fig.~\ref{model} summarises this picture of the circumnuclear environment of the AGN in NGC 7469.

%-------------------------------------------------------------
%
%_____________________________________________________________

%______________________________________________________________

\begin{acknowledgements}
      This work is based on observations obtained with XMM-Newton, an ESA science mission with instruments and contributions directly funded by ESA Member States and the USA (NASA). The MSSL authors acknowledge the support of PPARC. EB was supported by the Yigal-Alon Fellowship and by the GIF Foundation under grant \#2028-1093.7/2001. SRON is supported financially by NWO, the Netherlands Organization for Scientific Research.
\end{acknowledgements}

\end{document}